\documentclass[12pt]{iopart}
\usepackage{graphicx}
\begin{document}

\title{Quasi-Particle Density in Sr$_2$RuO$_4$ Probed by means of the Phonon Thermal Conductivity}

\author{M Suzuki$^1$, M A Tanatar$^{1,2}$\footnote[1]{Permanent address: Institute of Surface Chemistry, N.A.S. 
Ukraine, Kyiv, Ukraine; present address: Department of Physics, University of 
Toronto, Canada.}, Z Q Mao$^{1,2}$, Y Maeno$^{1,2,3}$\footnote[3]{Corresponding author(maeno@scphys.kyoto-u.ac.jp)} and T Ishiguro$^{1,2}$
}

\address{$^1$ Department of Physics, Kyoto University, Kyoto 606-8502, Japan \\
$^2$ CREST, Japan Science and Technology Corporation, Kawaguchi, Saitama 332-0012, Japan \\
$^3$ International Innovation Center, Kyoto University, Kyoto 606-8501, Japan \\}

\begin{abstract}
The thermal conductivity of Sr$_2$RuO$_4$ along the least conducting direction perpendicular to the RuO$_2$ plane has been studied down to 0.3 K.  
In this configuration the phonons remain the dominant heat carriers down to the lowest temperature, and their conductivity in the normal state is determined by the scattering on conduction electrons. 
We show that the phonon mean free path in the superconducting state is sensitive to the density of the quasi-particles in the bulk.  An unusual magnetic field dependence of the phonon thermal conductivity is ascribed to the anisotropic superconducting gap structure in Sr$_2$RuO$_4$.
\end{abstract}


%\submitto{\JPCM}

\maketitle

\section{Introduction}

Sr$_2$RuO$_4$ first attracted a lot of interest as a superconductor which has the layered perovskite structure without copper \cite{Nature}.  
Later on it has become clear that despite the crystal 
structural resemblance to the high-$T_{\rm c}\ $ cuprate superconductor La$_{2-x}$Sr$_x$CuO$_4$, 
the electronic properties of these two systems are quite different \cite{physicstoday}.  
The normal state of Sr$_2$RuO$_4$ is well described by Landau Fermi-liquid model
\cite{Mackenzie}, and its superconductivity is characterized by the spin 
triplet pairing and spontaneous breaking of the time-reversal 
symmetry \cite{Rice,Luke,Ishida1,Duffy}.

Study of the thermal conductivity can be very useful for probing the quasi-particle spectrum in the superconducting state \cite{review}. 
These measurements played an essential role in locating the position of 
the superconducting gap nodes in high-$T_{\rm c}\ $ cuprates \cite{YuAni,AubinAni} and in heavy fermion superconductors \cite{crystalanisotropy,IzawaCe}. Of note, in these systems the thermal conductivity is mostly determined by quasi-particle excitations and is electronic in origin, as confirmed in the cuprates by measurements of the Righi-Leduc effect (thermal Hall effect) \cite{ThermalHall}. Detailed experimental characterization of the electronic thermal conductivity stimulated intensive theoretical studies on the subject \cite{Sun,Graf,Barash,TheoryFD} and at present there is a reasonable understanding of the {\it electronic} thermal conductivity in unconventional superconductors.  
The thermal conductivity of Sr$_2$RuO$_4$ with the heat flow along the conducting plane was also shown to be electronic in origin \cite{SuderowRu,Tanatar}, and it is well described with a model for unconventional superconductors with line-nodes in the superconducting gap \cite{Tanatar,Izawa,Suzukiuniversal}. 

It is significant that Sr$_2$RuO$_4$ has a quasi-two-dimensional electronic 
structure \cite{Mackenzie,bergeman} and consequently, all of the electronic transport properties should be strongly anisotropic. 
Therefore it is of interest to extend the measurements to the configuration with the heat flow perpendicular to the plane. Here the low electronic conductivity provides a rare opportunity to study the {\it phonon} thermal conductivity in unconventional superconductors. 

In this article, we report measurements and analysis of the thermal 
conductivity along the least conducting direction (the [001] crystal axis) in the normal, superconducting and mixed states of Sr$_2$RuO$_4$.  
Through the analysis of the thermal conductivity in magnetic fields 
and the comparison of the results for the heat flow directions parallel and perpendicular to the plane, we separated 
the phononic and electronic contributions and characterized 
their field and temperature dependences. 

We revealed that the temperature and field dependence of the phonon 
thermal conductivity reflects an increased quasi-particle 
density in the bulk of unconventional superconductors.
A sub-linear field dependence of the phonon thermal resistivity 
is ascribed to the presence of nodes in the superconducting gap, which is consistent with other relevant experiments; specific heat \cite{Zaki}, NMR \cite{Ishida2}, magnetic penetration depth \cite{Bonalde} and ultrasonic attenuation \cite{Lupien}. 

\section{Experimental methods}

The single crystals of Sr$_2$RuO$_4$ used in this study were grown 
in an infrared image furnace by the floating zone method \cite{Mao}.  
In the crystal growth procedure, an excess Ru is charged to create the Ru self-flux in order to compensate for Ru metal evaporation 
from the surface of the crystal rod.  
Therefore, inclusions of Ru metal are frequently found in 
the central part of the crystal boule. Since these inclusions are responsible for the formation of the so-called 3K-phase \cite{3K}, to obtain the single-phase samples we cut them from 
the surface part of the boule. The absence of the 3K-phase was 
checked by the AC susceptibility measurements.  
We selected a high quality sample with $T_{\rm c}$ = 1.42 K, 
close to the extrapolated impurity - free value $T_{\rm c0}$ = 1.50 K \cite{defect,impurity}.  
The sample was cut into a rectangular slab elongated in the [001] direction 
with dimensions 1.5$\times$0.5$\times$0.5 mm$^3$.  

The thermal conductivity was measured by usual one heater - two thermometers 
method with three 2 k$\Omega$ RuO$_2$ resistance chips \cite{Resistor}.  The direction of the heat current was set along the [001] crystal axis perpendicular to the conducting plane. 
(Below for the thermal conductivity perpendicular to the plane we use the notation $\kappa_{\rm 001}$, with the subscript showing the direction of the heat flow.)
The end of the sample was attached directly to the cold finger of 
a miniature vacuum cell \cite{cell}.  
The cell was placed in a double-axis sample rotator 
in a $^3$He refrigerator inserted 
into a 17 Tesla superconducting magnet.  
The rotator enabled precise cell orientation with a relative accuracy of 
better than 0.1$^{\circ}$.  
Simultaneous electrical conductivity measurements were performed 
using the same thermal/electrical contacts.  

\section{Results and discussion}

\subsection{Temperature dependence of thermal conductivity} \label{temp_dep}

\begin{figure}
\begin{center}
\includegraphics[width=10cm,height=7cm]{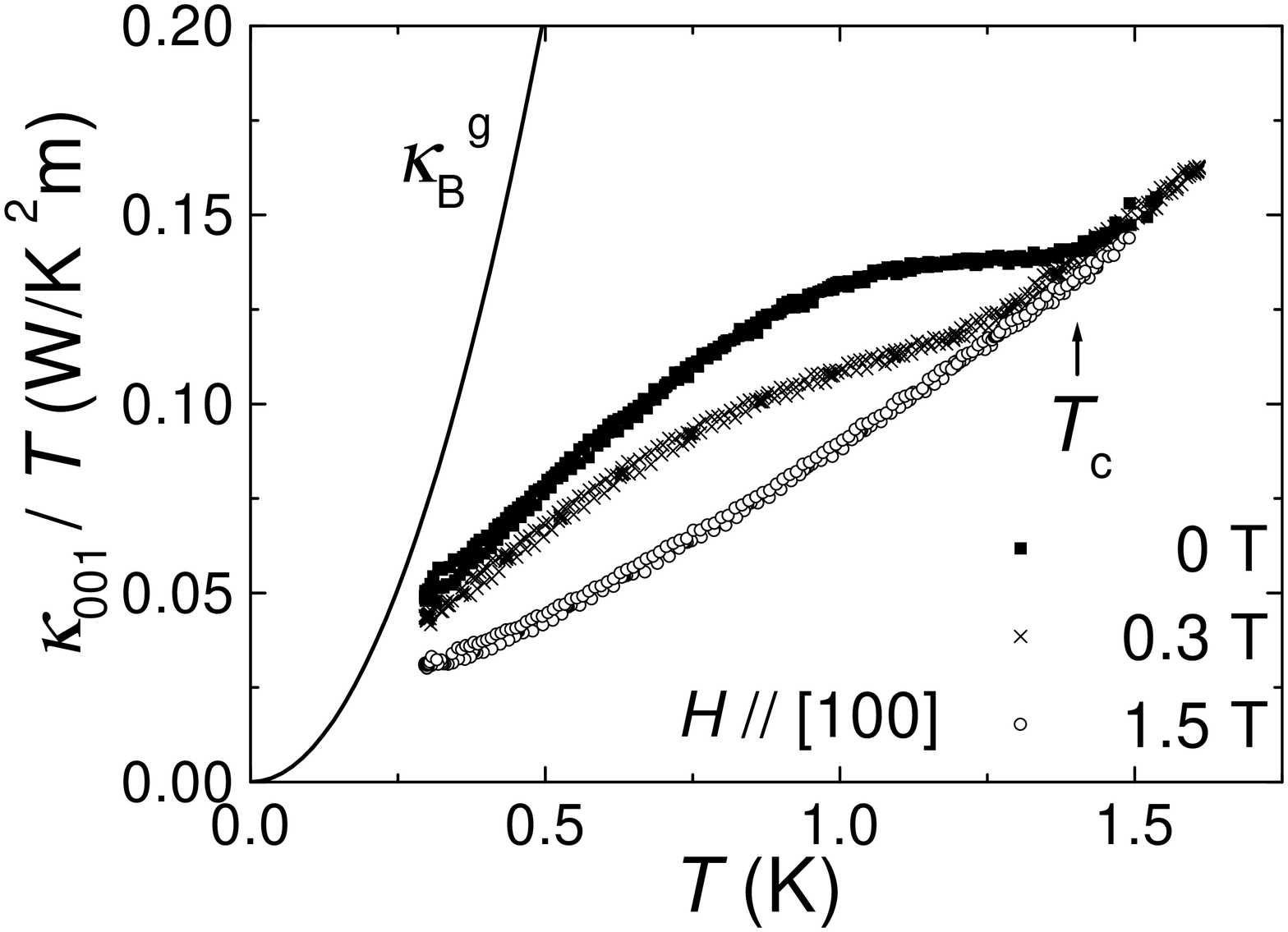}
\end{center}
\caption{
Temperature dependence of the thermal 
conductivity perpendicular to the highly conducting plane, $\kappa_{\rm 001}$,  of Sr$_2$RuO$_4$ in magnetic fields of 0 T, 0.3 T and 1.5 T 
(normal state) along [100] direction.  A solid curve is the calculated phonon thermal conductivity in the boundary scattering mode.  
}\label{td}
\end{figure}

In Fig. \ref{td} we show the temperature ($T$) dependence of 
$\kappa_{\rm 001}/T$ of ${\rm Sr_2RuO_4}$ in magnetic fields 
along the [100] direction.  $\kappa_{\rm 001}/T$ shows a shoulder on entering the 
superconducting state, contrary to the in-plane thermal 
conductivity $\kappa_{\rm 100}$, of the same high quality ($T_{\rm c}$ = 1.44 K) crystals, which shows an immediate decrease below $T_{\rm c}$ \cite{Tanatar}.  

To obtain an appropriate understanding 
of these results, the identification of the dominant heat carriers and the scattering mechanisms is important.  
The thermal conductivity $\kappa$ is a sum of contributions of electrons and phonons, $\kappa = \kappa^e + \kappa^g$, and in the 
superconducting state both of these quantities change in a complicated way.  
In the normal state, however, throughout the whole temperature range studied we are already in the residual resistivity range, therefore, the electronic component $\kappa^e/T$ should be constant with $T$. 

The simplest way to estimate $\kappa^e$ is to calculate it from the value of the electrical resistivity $\rho$. The two quantities are related by the 
Wiedemann-Franz (WF) law, $\rho \times (\kappa^e/T) = L_0$, 
where $L_0$ is a Lorenz number of $2.45 \times 10^{-8} {\rm W\Omega/K^2}$.  
This law was shown to be satisfied in Sr$_2$RuO$_4$ \cite{Tanatar}, contrary to the electron-doped cuprate superconductors \cite{Hill}. To exclude large errors in determination of the sample geometry, 
we compare the measurements of both quantities made on 
the same electrical/thermal contacts. Using the resistance value in the field of 1.5 T along [100] direction, the electronic contribution to $\kappa_{\rm 001}$ in the normal 
state is evaluated as $\kappa_{\rm 001}^{e}/T$ = $9.9\ {\rm mW/K^2m}$, 
which amounts to 7\% of the total $\kappa_{\rm 001}$ in 
the normal state at 1.5 K and 32\% at 0.3 K.  The validity of the WF law is based on the hypothesis that the electronic 
conduction is determined by the elastic scattering process. 
The inelastic processes reduce the thermal current more effectively than the 
charge current, giving the WF ratio smaller than $L_0$.  Thus this method of estimation provides the upper bound of the actual electronic contribution. 

Alternative way of estimation of $\kappa^{e}$ uses the assumption that relative changes of electrical conductivity $\rho^{-1}$ and of $\kappa^e$ with fields are the same \cite{Roeske}. In this way, using the data for magnetothermal resistance and magnetoresistance in the normal state with field along [100] direction, we obtain a consistent value $\kappa_{\rm 001}^{e}/T$ = $9.5\ {\rm mW/K^2m}$, 7\% of measured $\kappa_{\rm 001}/T$ at 1.5 K and 30\% at 0.3 K.  

As we see above, in the normal state, the thermal conductivity is mainly phononic in origin. This is moreover true for the superconducting state, since here, due to the dominance of impurity scattering for conduction electrons, the electronic contribution $\kappa^e/T$ shows notable additional decrease as compared to the normal state \cite{Tanatar}. Therefore we can conclude that the 
the enhancement of $\kappa_{\rm 001}$ below $T_{\rm c}$ is due to the phonon mean free path 
increase, the cause of which will be discussed below.  

Possible scattering mechanisms in the phonon heat conduction include the scattering on the sample boundaries, conduction electrons in the normal state, thermally excited quasi-particles and vortex cores in the superconducting state.  As can be seen in Fig. \ref{td}, above $T_{\rm c}$, $\kappa_{\rm 001}/T$ 
changes linearly with temperature ($\kappa_{\rm 001}/T \propto T$),
which shows that the $\kappa^g$ is limited mainly by scattering on conduction 
electrons.  
Thus the hump of $\kappa_{\rm 001}/T$ observed below 
$T_{\rm c}$ is interpreted as a result of the condensation of electrons into 
the Cooper pairs which do not contribute to the scattering of phonons.  For the conventional superconductors, Bardeen, Rickayzen and Tewordt (BRT) 
calculated the increase of $\kappa^g$ due to the enhancement of the phonon mean free path caused by the formation of the Cooper pairs 
below $T_{\rm c}$ \cite{BRT}, assuming an exponential decrease of 
the quasi-particle density.  According to their theory, almost 10 times increase of $\kappa^g$ is expected at the temperature of 0.5 $T_{\rm c}$ \cite{BRT}.  It can be seen that the actual increase of the thermal conductivity is far 
lower than this expectation, and the large difference indicates the existence of additional scatterers 
such as grain/sample boundaries or excess quasi-particles resulting from the existence of nodes in the gap. 

The role of the boundary scattering can be evaluated in the following way. 
In the boundary scattering regime, the phonon thermal conductivity is described by the asymptotic formula $\kappa^g_B = \frac{1}{3}C\bar{v}l \propto T^3$, 
where $C$=$\beta T^3$ is the phonon specific heat per unit volume, $\bar{v}$ 
is the mean sound velocity and $l$ is the mean free path of the heat carriers 
limited by the sample dimension.  The mean sound velocity along the [001] direction of the tetragonal crystal is 
described by the relation $\bar{v} = v_L\frac{2s^2+1}{2s^3+1}$, 
where $s = v_L/v_T$ is the ratio of the longitudinal and transverse sound 
velocities \cite{Carruthers}. The mean free path of phonons in the boundary scattering regime is 
$l_B = \frac{2d}{\sqrt{\pi}}$, where $d$ is the mean width of the  sample cross section \cite{Carruthers}.  
If we take a phonon specific heat coefficient of $\beta$= 0.197 
${\rm mJ/K^4 mole}$ \cite{Zaki}, sound velocities $v_L = 3900~{\rm m/s}$ 
and $v_T = 1300~{\rm m/s}$ at low temperatures \cite{velocity} and $d = 0.5~{\rm mm}$, we obtain the phonon thermal conductivity in the boundary scattering regime as $\kappa^g_B$ = 0.8 $T^3$ W/Km.  This curve is plotted in Fig. \ref{td} with a solid line, which lies well above the actual data in the whole temperature range.  This means that the phonon thermal conductivity is strongly limited by the large amount of the thermally excited quasi-particles in the vicinity of nodes.  Actually the observed height of the hump below $T_{\rm c}$ is 8 times lower than the prediction of the scattering on the sample boundaries.  

\subsection{Field dependence of thermal conductivity}

Now we discuss the field dependence of the phonon thermal conductivity.  For the separation of the $\kappa^e$ and $\kappa^g$, we analyze the different field dependence expected for two kinds of heat carriers in comparison with the known field dependence for $\kappa^e$ \cite{Tanatar}.  

\begin{figure}
\begin{center}
\includegraphics[width=9cm,height=13cm]{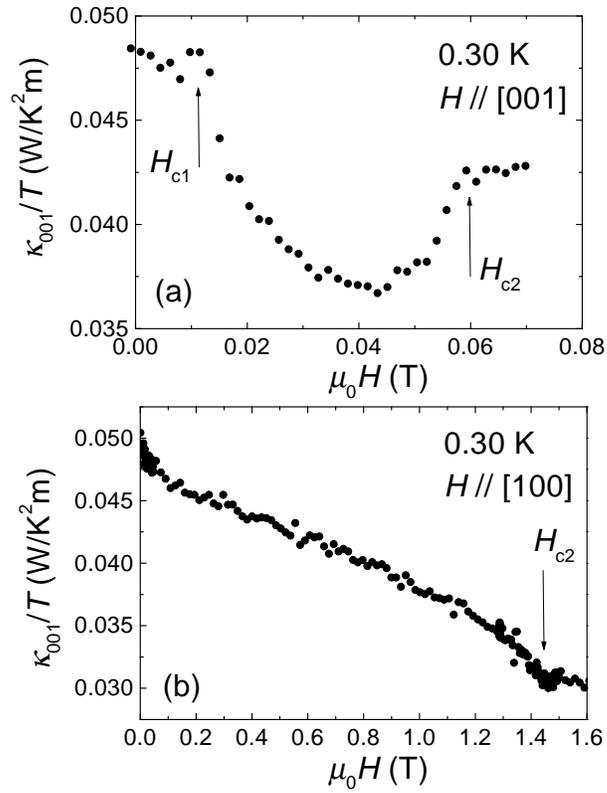}
\end{center}
\caption{The magnetic field dependence of $\kappa_{\rm 001}/T$ at 0.30 K; (a) in magnetic fields along the [001] direction. (b) in magnetic fields along the [100] direction.
}\label{fd}
\end{figure}

The field dependence of $\kappa_{\rm 001}/T$ at 0.3 K in the magnetic  
field applied along the [001] direction is shown in Fig. \ref{fd} (a).  On field increase from 0 T $\kappa_{\rm 001}/T$ 
first remains nearly constant, then shows a sharp drop, takes a minimum and starts to increase rapidly towards $H_{\rm c2}$.  Finally $\kappa_{\rm 001}/T$  flattens above $H_{\rm c2}\ $ in the normal state.  The plateau observed at low fields corresponds to the complete Meissner state with $H_{\rm c1}$ $\sim$ 12 mT at 0.3 K \cite{Hc1}.  

\begin{figure}
\begin{center}
\includegraphics[width=10cm,height=7cm]{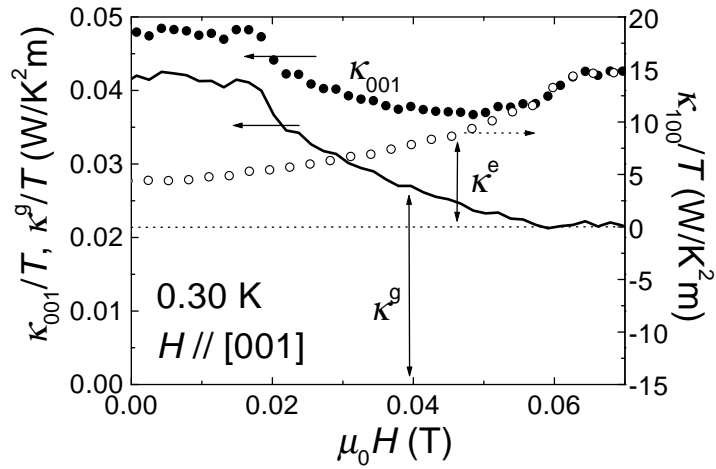}
\end{center}
\caption{The magnetic field dependence of $\kappa_{\rm 001}/T$  in magnetic fields along the [001] direction at 0.30 K (closed symbols). Open symbols represent the field dependence of the thermal conductivity with the heat flow along the conducting plane $\kappa_{100}$, measured in the same field orientation and temperature, scaled to fit the increase of $\kappa_{\rm 001}$ near $H_{\rm c2}\ $. 
The solid curve shows the difference, representing the field dependence of 
the phonon conductivity $\kappa_{001}^{g}$. 
}\label{sub}
\end{figure}

Since the effect of magnetic fields on the phonon thermal conductivity comes only from the creation of additional scatterers, $\kappa^g$ decreases monotonically with fields. The field dependence of $\kappa^e$ of Sr$_2$RuO$_4$ is more complicated. At high temperature $\kappa^e$ shows initial decrease with fields, and then increases towards $H_{\rm c2}$ \cite{Tanatar}.  However, at low temperatures $\kappa^e$ increases monotonically with fields \cite{Tanatar}. Thus the rapid increase of $\kappa_{\rm 001}/T$ near $H_{\rm c2}$ comes from the increase of $\kappa^e$ on the slowly decreasing background of $\kappa^g$.  This allows us to decompose the field dependence into electronic and phononic components.  
We assume that the electronic part in $\kappa_{\rm 001}$ has the same field dependence, except for difference in amplitude,  with as $\kappa_{\rm 100}$ which is purely electronic in origin. Then the field dependence of $\kappa^g_{\rm 001}$ can be determined by subtracting the scaled $\kappa_{\rm 100}$ (i.e. $\kappa^e_{\rm 001}$) from the total $\kappa_{\rm 001}$ as shown in Fig. \ref{sub}.  Here the scaling factor is used as adjustable parameter, chosen so as to give coincidence of the scaled $\kappa_{\rm 100}$ and $\kappa_{\rm 001}$ curves near $H_{\rm c2}\ $.  
As shown in the figure, in the superconducting state, the phonon 
contribution is strongly dominant, accounting for 88\% of $\kappa_{\rm 001}/T$ 
at 0~T and 0.3~K. Although being efficient, this procedure is, strictly speaking, not correct because we neglected the difference in magnetoresistance and hence magnetothermal resistance between $\kappa^e_{\rm 100}$ and $\kappa^e_{\rm 001}$. However, since the magnetic field necessary to suppress the superconducting state in $H \parallel [001]$ configuration is small, this correction is not of large importance here. 

On the contrary, in the magnetic fields parallel to the plane, the magnetoresistance becomes large for the current perpendicular to the plane even just above $H_{\rm c2}$ \cite{Ohmichi}. In our case the resistance at $H_{\rm c2}$ in perpendicular field is approximately 1.7 times smaller than in parallel field. This magnetoresistance accounts for the smaller normal state electronic thermal conductivity in parallel fields (30-32~\% of the total conductivity at 0.3 K, see Section \ref{temp_dep}) as compared to perpendicular field case, in which case $\kappa^e$ constitutes about a half of the total normal state conductivity (Fig. \ref{sub}).  

The dependence of $\kappa_{\rm 001}/T$ on magnetic fields $H \parallel$ [100] (along the plane) at 0.3 K is shown in Fig. \ref{fd} (b).   Contrary to the perpendicular field case, $\kappa_{\rm 001}/T$ curve does not have the minimum, resulting from increase of the electronic conductivity towards $H_{\rm c2}$. Instead, the conductivity decreases gradually and then shows a rapid {\it decrease} towards the normal state value in a narrow field range in the vicinity of $H_{\rm c2}$.  
We point out here, that the absence of the upturn at $H_{\rm c2}\ $ in parallel fields at low temperatures, despite a notable increase of $\kappa^e$ \cite{Tanatar}, is completely unusual. This fact implies that (i) phonon scattering is strongly suppressed in the superconducting state at low fields, (ii) some additional mechanism for phonon scattering appears close to $H_{\rm c2}$, which overcomes a rapid increase of $\kappa^e$. The latter fact cannot find any explanation in the standard vortex scattering scenario.  As is clear from the comparison with Fig. \ref{sub}, the absence of the upturn rules out the application of the procedure we used in the perpendicular field for extracting $\kappa^g (H)$ in the superconducting state.

As we already mentioned above, the field dependence of the phonon thermal conductivity is totally determined by the variation of the scattering rate. The scattering in the superconductor is governed by the vortex cores and delocalized quasi-particles in the bulk, the latter acting similar to conduction electrons in the normal state.  For a superconductor with an anisotropic gap with line nodes the quasi-particles in the bulk are generated by a magnetic field, as was first pointed out by Volovik \cite{Volovik}. This effect is caused by a supercurrent flow around a vortex and associated local Doppler shift of the quasi-particle spectrum. The shift causes influx of the quasi-particles from the core into the bulk along the directions of the nodes in the superconducting gap.  This increase of the quasi-particle density in fields is indeed observed in specific heat measurements of Sr$_2$RuO$_4$ \cite{Zaki}, giving a strong evidence for the nodal structure of the superconducting gap. 
At low temperatures the density of delocalized QP is negligible in conventional type-II superconductors with the isotropic gap,  therefore the phonon scattering in magnetic fields between $H_{\rm c1}$ and $H_{\rm c2}$ proceeds predominantly on the vortex cores. In this case the phonon thermal resistivity $W(H)$ increases linearly with $H$ for $H>H_{\rm c1}$, being proportional to the number of vortices. It can be expressed \cite{Lowell} as
\begin{eqnarray}
W(H)\ =\ \frac{1}{\kappa^{g}}\ =\ W_{\rm 0}+W_{\rm n}H/H_{\rm c2}, \label{eq1}
\end{eqnarray}
where $W_0$ and $W_0+W_{\rm n}$ represent the phonon thermal resistivity in the Meissner and normal states, respectively.  

\begin{figure}
\begin{center}
\includegraphics[width=10cm,height=7cm]{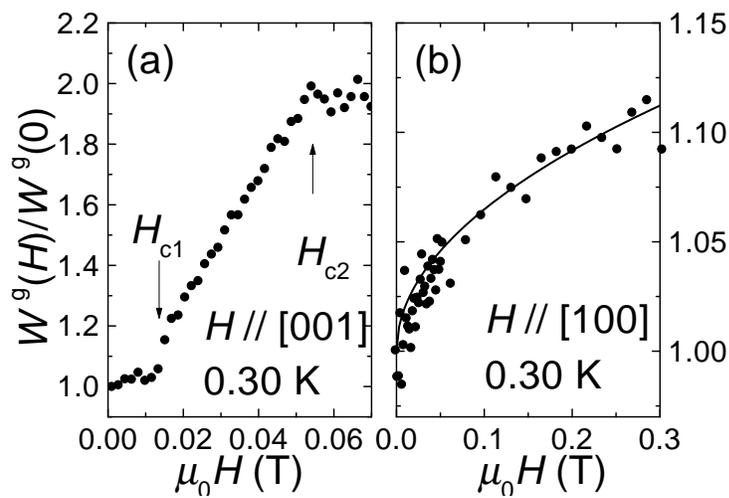}
\end{center}
\caption{(a) The dependence of the phonon thermal resistivity on magnetic fields perpendicular to the plane ($H \parallel$ [001]), extracted in the way shown in Fig. \ref{sub}.  
(b) The low field part of the field dependence of the thermal resistivity in magnetic fields along [100] direction within the plane (symbols).  The line represents a least square fit of the data with a $\sqrt{H}$-function.  
}\label{W}
\end{figure}

Fig. \ref{W} (a) shows the field dependence of the phonon thermal resistivity denoted as $W^g(H)$ in the magnetic field $H \parallel$ [001].  The phonon part was extracted as shown in Fig. \ref{sub}. $W^g(H)$ increases linearly with field and shows the dominance of the vortex core scattering. In this field configuration the coherence length is large ($\xi_{ab} \sim 660~{\rm \AA}$)\cite{Akima}, giving a large cross-section of the vortex core for the phonon scattering even at low fields.  Thus the delocalized quasi-particles, despite their density is large as seen in the specific heat measurement \cite{Zaki}, do not effectively influence the phonon transport.  

In Fig. \ref{W} (b) we plot the thermal resistivity $W(H)$ at 0.3 K as a function of magnetic fields parallel to the plane ($H \parallel$ [100]).  Here $W(H)$ is obtained directly from 1/$\kappa_{001}$ and consequently contains a small electronic contribution.  However, for fields $H<H_{\rm c2}/2$, the electronic contribution is slightly varying with fields \cite{Tanatar}, and should give small and smooth offset of zero line, hence the {\it shape} of $W(H)$ can well represent that of $W^g(H)$. As it can be seen from the figure, the $W(H)$ shows a notable sublinearity and at low fields exceeds the $H$-linear increase expected for scattering on the vortex lattice. This shows that the vortex scattering is not dominant in this configuration, which can be understood if we recall that the vortex core for parallel field is approximately 20 times smaller ($\xi_{c} \sim 33~{\rm \AA}$)\cite{Akima} than in the perpendicular field. In the situation when the vortex scattering is weak, the delocalized quasi-particles induced by magnetic fields can become the main scatterers of phonons.  

A quantitative description of the role of the delocalized quasi-particles on the phonon transport is very difficult at present since it is almost unexplored theoretically.  As a naive discussion, however, we can assume that the thermal resistivity is proportional to the number of scatterers, i.e. the delocalized quasi-particles.  This density was calculated by Volovik for isolated single vortex core in the $d$-wave superconductor \cite{Volovik} and it is proportional to $\sqrt{H}$ at very low temperatures. Thus we may consider that the phonon resistivity should be proportional to $\sqrt{H}$.  The curve in Fig. \ref{W} (b) is a most close fit of the data with a $\sqrt{H}$-function.  While this fit reproduces our result well, the above discussion is oversimplified and a proper theory for the phonon transport in the unconventional superconductor is strongly desired.  Very recently, Won and Maki \cite{WonMaki} calculated the field dependence of the phonon thermal resistivity in the nodal superconductor, assuming dominance of the longitudinal phonons in the heat transport at low temperatures and the validity of the Matthiesen rule for rates of scattering on sample boundaries and on quasi-particles. They derived a sub-linear field dependence of the form $HlnH$ due to the delocalized quasi-particles \cite{WonMaki}, which is not very distinct from $sqrt(H)$ dependence at low fields and seems consistent with our results within the experimental error.  

\begin{figure}
\begin{center}
\includegraphics[width=10cm,height=7cm]{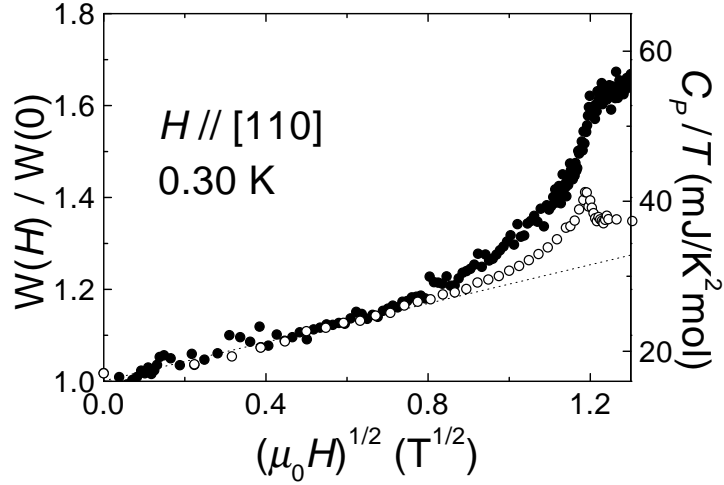}
\end{center}
\caption{The field dependence of the thermal resistivity $W(H)$ (closed symbols) and specific heat (open symbols) in Ref. [23] in magnetic fields along [110] direction within the plane at 0.30 K.  The horizontal axis is scaled as $\sqrt{H}$.  
}\label{specific}
\end{figure}

To check the relation between the phonon resistivity and the quasi-particle density, we compare in Fig. \ref{specific} the field dependence of the thermal resistivity and of the specific heat by Nishizaki {\it et al}. \cite{Zaki} representing the quasi-particle density, both taken at 0.3 K in the same experimental configuration with in-plane fields. Note that the horizontal axis is scaled as $\sqrt{H}$.  It can be seen that at low fields both quantities show similar sublinear dependence and match well, supporting our assumption on the phonon scattering on the delocalized quasi-particles. Despite this similarity, the two curves can never coincide completely, since corrections due to the electronic contribution and additional scattering on the vortex lattice are essential for the thermal conductivity at high fields.  
When scaled to overlap in the low field region, the thermal resistivity appears to be higher than the specific heat curve.  
The electronic contribution $\kappa^{e}$, which can become important near 
$H_{\rm c2}$, should cause the decrease of the resistivity and thus can not be 
the reason for this deviation. Therefore we conclude that additional scattering on vortex lattice is important as well. 

An additional support for the role of scattering on field-induced quasi-particles in Sr$_2$RuO$_4$ in parallel field comes from an unusual shape of the $\kappa(H)$ curve near $H_{\rm c2}$ at low temperatures. As can be deduced from Eq. (1), in fields close to $H_{\rm c2}$ the decrease of the phonon contribution due to scattering on the vortex lattice saturates to a value equal to that in the normal state. Therefore, as in the perpendicular field case, the increase of the $\kappa^e$ towards $H_{\rm c2}$ over this flat background gives an increase of total $\kappa$. However, the situation in Sr$_2$RuO$_4$ in magnetic fields parallel to the plane is notably different. Due to a multi-component order parameter, the superconducting state of Sr$_2$RuO$_4$ is characterized by existence of at least two superconducting phases \cite{Agterberg}. The main superconducting state occupies almost all $H-T$ domain of the superconductivity, while the second superconducting state exists at low temperatures (below 0.8 K) at precisely aligned (within $\pm 3^{\circ}$ close to the plane) fields in close proximity to $H_{\rm c2}$  \cite{Tanatar,Zaki,Yaguchi}. 

A phase transition from low field superconducting phase to high-field one is accompanied by a release of large density of quasi-particles. This release of QP gives a rapid increase of electronic thermal conductivity, which in the normal situation should give an upturn in $\kappa(H)$. However, since the release of the quasi-particles is similarly important for the phonon scattering, and the phonon contribution $\kappa^g$ is much larger than $\kappa^e$, the decrease of $\kappa^g$ is dominating over increase of $\kappa^e$ and gives downturn in $\kappa_{001}(H)$. 

\begin{figure}
\begin{center}
\includegraphics[width=10cm,height=7cm]{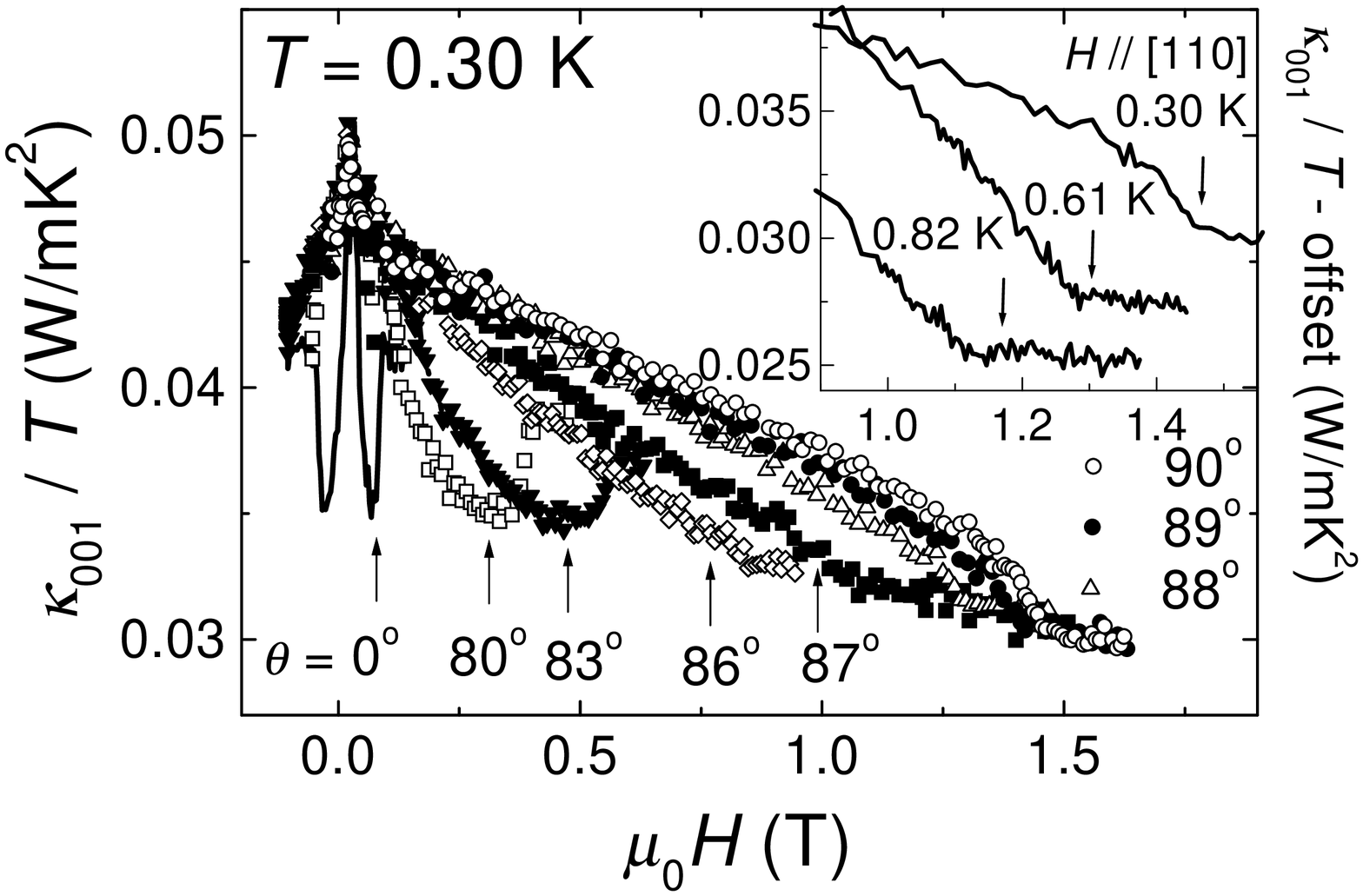}
\end{center}
\caption{Transformation of the field dependence of $\kappa_{\rm 001}/T$ at base temperature (0.30 K) with direction of the field. The numbers show an angle of field inclination to the conducting plane, being 90$^{\circ}$ for parallel alignment; the [110] direction. Inset; the magnetic field dependence of $\kappa_{\rm 001}/T$ at several temperatures with $H$ applied parallel to the conducting plane ([110]).  Note that some offset is subtracted by each curve.  Arrows show $H_{\rm c2}$ for each curve. 
}\label{inclination}
\end{figure}

To see whether the unusual curvature is indeed related to the formation of the second phase we show in the inset of Fig. \ref{inclination} the $\kappa(H)$ dependence at 0.82 K, above the domain of the second superconducting phase. As can be seen, the curve shows usual shape with a small upturn in $\kappa(H)$ near $H_{\rm c2}$, despite notably smaller electronic contribution to the total thermal conductivity at this elevated temperature. Simultaneously, as shown in the main panel of Fig. \ref{inclination}, the downward curvature appears in the same range of field inclinations (within $\pm3^{\circ}$ to the plane) where the second superconducting state exists and where rapid release of quasi-particles is observed in specific heat and electronic thermal conductivity measurements \cite{Tanatar,Zaki}.  These observations clearly show that the downturn is determined by formation of a second superconducting phase, and is determined by a rapid increase of scattering on quasi-particles near $H_{\rm c2}$. 

\section{Conclusion}

The thermal conductivity perpendicular to the conducting plane of Sr$_2$RuO$_4$ down to the lowest temperatures is dominated by phonon contribution, with scattering by conduction electrons or quasi-particles as the dominant scattering mechanism. The shoulder in the temperature dependence of the thermal conductivity in the superconducting state is related to the enhancement of the phonon mean free path due to the condensation of normal electrons below the superconducting gap, and its height is substantially lower than the expectation for a superconductor with a nodeless gap. In the mixed state the phonon heat transport is determined by the scattering on the vortex cores and on the bulk quasi-particles outside the core. The latter contribution is important in the configuration when the magnetic field is parallel to the plane, where the core radius is small.  These results imply that the phonon thermal conductivity is a useful probe for the characterization of the unconventional superconducting state, complementary to the electronic thermal conductivity or specific heat \cite{PRL}.

\vspace{5mm}
{\bf Acknowledgement}
\vspace{5mm}

The authors thank A E Kovalev and V A Bondarenko for 
their contribution to the experiment design.  They are grateful to S Nishizaki for providing specific heat data and to H Won and K Maki for presenting the calculation results prior to publication. The authors acknowledge A G Lebed, Y Matsuda and M Sigrist for invaluable comments and discussions.  This work was partly supported by the Grant-in-Aid for Scientific Research on Priority Area "Novel Quantum Phenomena in Transition Metal Oxides" from the Ministry of Education, Culture, Sports, Science and Technology. 

%%%%%%%%%%%%%%%%%%%%%%%%%%%%%%%%%%%%%%%%%%%%%%%%%%%%%%%%%%%%%%

%%%%%%%%%%%%%%%%%%%%%%%%%%%%%%%%%%%%%%%%%%%

\end{document}